\documentstyle[equations,epsf]{mn}
\makeatletter
\def\@cite#1#2{(\if@tempswa #2 \fi #1)} 
\makeatother
\def\rd{{\rm d}}                                      
\def\rdif#1#2{\mathchoice{\rd{#1}\over\rd{#2}}{\rd{#1}/\rd{#2}}
{\rd{#1}/\rd{#2}}{\rd{#1}/\rd{#2}}}                            
\def\refeq#1{{(\ref{#1})}}
\newcommand{\bv}[1]{\mbox{$\bf #1$}}
\def\unit#1{\,{\rm {#1}}}
\def\umult#1#2{\ifx#2\unit\def\tmpa{\umulti#1#2}\else\toks0=\expandafter{#2}%
\edef\tmpa{\noexpand\umulti\noexpand#1\the\toks0}\fi\tmpa}
\def\umulti#1#2#3{\ifx#2\unit\unit{#1#3}\else%
\message{\noexpand\umult error: must precede unit or \noexpand\unit}%
\unit{#1}#2#3\fi}

\def\secnd{\unit{s}}

\def\cm{\unit{cm}}

\def\parsec{\unit{pc}}

\def\erg{\unit{erg}}

\def\Kelv{\unit{K}}

\def\subrm#1{_{\rm #1}}
\catcode`|=\active \let|=\subrm
\def\ee{\protect\pee}
\def\pee#1{\ifmmode{\times10^{#1}}\else$\times10^{#1}$\fi}
\def\ion#1#2{{\rm #1}%
\ifmmode{\mathchoice{\scriptstyle}{\scriptstyle}
{\scriptscriptstyle}{\scriptscriptstyle}{\rm\uppercase{#2}}}%
\else$\,\scriptstyle\rm\uppercase{#2}$\fi}
\def\HI{{\ion{H}{I}}}
\def\HII{{\ion{H}{II}}}

\def\eg{{e.g.}}
\def\ie{{i.e.}}

\def\cf{{cf.}}
\title[Recombination fronts]
{Breaking the sound barrier in recombination fronts}
\author[R.J.R. Williams \& J.E. Dyson]{R.J.R. Williams \& J.E. Dyson\\
Department~of~Physics~and~Astronomy,
University~of~Manchester, Oxford~Road, Manchester M13~9PL}
\date{Received **INSERT**; in original form **INSERT**}
\pagerange{\pageref{firstpage}--\pageref{lastpage}}

\begin{document}
\label{firstpage}
\maketitle

\begin{abstract}
We exploit a generic instability in the integration of steady, sonic,
near-isothermal flows to find the complete transition diagram for
recombination fronts (for a model system of equations).  The
instability requires the integration of the flow equations for speeds
between the isothermal and adiabatic sound speeds to be performed with
particular care.  As a result of this, the previous work of Newman \&
Axford on the structure of recombination fronts neglected an important
class of solution, that of transonic fronts; our method is readily
extensible to a more complete treatment of the ionization structure.
Future papers will apply these results in models of the structure of
ultracompact \HII\ regions.
\end{abstract}


\section{Introduction}

The hydrodynamics of gas photoionized by a central star has been
widely studied \cite[\eg{}]{kahn54,axfo61,gold61}.  The gas is almost
completely ionized close to the central source.  The incident
radiation field decreases with radius as a result of geometrical
divergence and because atomic recombination always results in a small
population of neutral atoms, which have finite opacity to the ionizing
radiation.  At some radius the gas becomes optically thick to the
ionizing radiation, and beyond this radius it is predominantly
neutral.

The transition between ionized and neutral gas can be treated as a
quasi-steady, plane-parallel front when it takes place over a small
distance, compared with the radius of the region, and can be included
in models of the global flow as a discontinuity \cite[\eg{}]{oste89}.
This is, in general, the appropriate case for ionization fronts (IF),
since the opacity of the gas to ionizing radiation increases rapidly
as a function of decreasing ionization.  In this paper, we will
distinguish recombination fronts (RF) from IF as the case when the
net flow of gas is from high ionization towards low ionization.  RF
can be broadened significantly by the advection of ionized gas,
whereas IF are generally only as thick as the (small) photon mean free
path in neutral gas.

Newman~\& Axford~(1968 -- henceforth NA) studied the possible RF
transitions in the context of stellar winds.  They found that thin
fronts were unlikely to occur in the winds of solar-type stars but
might be important in the ejecta of planetary nebulae.  In recent
studies of ultracompact \HII\ regions \cite{dyso94,wild94,dywr95}, we
have discussed the likely importance of (thin) RF in the flow
structures.  In particular, the likely coincidence of RF and sonic
transitions raised some questions with regard to the work of NA, which
we address in this paper.

The significance of the small class of steady solutions which pass
through sonic transitions at the RF is perhaps not immediately
apparent when only conditions close to the front are considered.
However, such solutions become important when the front is included in
the global solution of an initially subsonic divergent flow.  Here,
the ballistic character of the supersonic solutions allows such
`winds' to satisfy a wide range of external pressure boundary
conditions, by the inclusion of a shock at some radius, for external
pressures far lower than required if the solutions are subsonic at all
radii \cite{park58}.

As part of this study, we also address the apparent paradox that for
conditions arbitrarily close to isothermal, the flow equations are
singular at the {\it adiabatic}\/ sound speed, whereas in the
isothermal limit they are singular at the {\it isothermal}\/ sound
speed.

In the following sections, we give simple model equations for steady
ionized gas flow, and derive a second order system which describes the
structure of plane IF and RF.  A reorganisation of these equations
suggests that for a narrow range of flow speeds in the region where
the flow is nearly isothermal, a formal instability will prevent
direct integration of the Mach number equation in the downstream
direction; this is illustrated by a yet simpler model system.  With
this in mind, we return to RF, and find a range of transonic solutions
which were inadvertently missed by NA.  In conclusion, we briefly
discuss the relevance of these results to models of ultra compact
\HII\ regions, and note possible implications of our study for the
details of the evolution of conventional IF in \HII\ regions at the
time of shock generation.

\section{Basic equations}

Differential equations for the structure of ionization and
recombination fronts are given by NA \cite[\cf{} also]{frnb92}.  Terms
which account for photoionization and recombination, the radiative
transfer of ionizing radiation and the heating and cooling of the gas
are added to the basic equations of hydrodynamics.  The processes
included are ionization of atomic hydrogen, proton and electron
recombination in the on-the-spot approximation, and an
approximation to forbidden line cooling ($L|{ei}$).
Mason~\shortcite{maso77} gives far more detailed models of the
internal structure of several types of IF, showing, for example, the
ionization structure of O, N and Ne.  For the present paper, however,
the approximations of NA will suffice.

The following equations give a reasonable approximation to the
structure of RF, in the frame in which the front is steady,
\cite[\cf{}]{axfo61,newa68}:
\begin{subequations}
\begin{eqnarray}
\nabla.(\rho\bv{v}) & = & 0 \label{e:mass}\\
\nabla.([\rho \bv{v}]\bv{v}) & = & -\nabla p \label{e:momm}\\
\nabla.\left(\left[w+{1\over2}\rho v^2\right]\bv{v}\right) & = &
	\rho Q\\
\nabla.([\rho x]\bv{v}) &=&
	\rho\left[(\alpha J)(1-x)-\beta n x^2\right]
	\label{e:fioniz}\\
\nabla.\bv{J} & = & -n(\alpha J)(1-x),
\end{eqnarray}%
\label{e:hydroabs}%
\end{subequations}%
where $\rho$ is the mass-density in the gas, $\bv{v}$ is the velocity,
$p$ is the pressure and $x$ is the fractional ionization.  The equation
of state,
\begin{equation}
p = (1+x)nkT,
\end{equation}
relates these quantities to $n = \rho/m|H$, the nucleon number
density, and the gas temperature $T$. The photoionization
cross-section of hydrogen at the Lyman limit is taken to be $\alpha =
6.3\ee{-18}\cm^2$ per atom, and $\beta$, the case B recombination
coefficient, is $\beta_0 T^{-3/4}$, where we use the value $\beta_0 =
3\ee{-10}\cm^3\secnd^{-1}\Kelv^{3/4}$ from NA.

The internal energy per unit volume is $\epsilon = p/(\gamma-1)$; the
enthalpy per unit volume is $w = \epsilon + p$.  $\bv{J} \equiv
J\hat{\bv{r}}$ is the outward flux of ionizing photons.

$Q$ is an effective heating rate, given by
\begin{equation}
\rho Q = k T_\star n (\alpha J) (1-x) - {1\over\gamma-1}\beta k T n^2 x^2
	-n^2x^2L|{ei},\label{e:heat}
\end{equation}
where $T_\star$ is the effective stellar temperature (which we take to
be $40\,000\Kelv$).  The final term in equation~\refeq{e:heat} is a
fit to forbidden line cooling \cite{gold61}, where
\begin{equation}
L|{ei} = \left\{
	\begin{array}{lc}
	L|{ei0}(T-4000\Kelv)^2 & \mbox{for $T>4000\Kelv$}\\
	0			& \mbox{otherwise}
	\end{array}\right.
\end{equation}
and $L|{ei0} = 9.7\ee{-32}\erg\cm^3\secnd^{-1}\Kelv^{-2}$.  This fit
is poor for gas temperatures above $10^5\Kelv$, but since temperatures
as high as this are not found in the fronts that we are studying, and
since we wish to compare our results to those of NA, we defer a more
accurate treatment of forbidden line cooling to the future.

Boundary conditions (at small $r$) are of high ionization ($x\to 1$),
high ionizing flux ($J\to\infty$), equilibrium temperature ($Q\to 0$)
and an arbitrary initial velocity ($v > 0$).  These are, however, not
sufficient to fully determine the flow in all cases, as will be
described below.

\section{Steady state solutions}

For a thin front, the mass flux and momentum equations (\ref{e:mass},
\ref{e:momm}) can be integrated immediately.  The remaining ordinary
differential equations which describe the front are
\begin{subequations}
\begin{eqnarray}
\rdif{m}{r} &=& {1+\gamma m^2\over 1-m^2} \left((\gamma-1)Q\over 2c^3\right)
	\label{e:mfront} \\
\rdif{x}{r} &=& n\left( \alpha j (1-x) - {\beta x^2\over v} \right) \\
\rdif{j}{r} &=& -n\alpha j(1-x),
\end{eqnarray}%
\label{es:thinfront}
\end{subequations}%
where $c$ is the adiabatic sound speed, $c^2 =\gamma p/\rho$, $m$ is
the {\it adiabatic} Mach number, and $j\equiv J/nv$ is the
dimensionless ionizing flux.  The integrals of the mass and momentum
flux equations, $\Phi$ and $\Pi$ respectively, are
\begin{equation}
\Phi = \rho v; \quad
	\Pi = \left[{1\over\gamma} + m^2\right]\rho c^2.\label{e:aux}
\end{equation}
Hence $\Pi/\Phi = (1+\gamma m^2)c/\gamma m$.  In this system of
equations, the radial distance scales simply with density, and so we
can study two coupled ODEs for $\rdif{m}{j}$ and $\rdif{x}{j}$ which
are equivalent to the equations studied by NA.

\subsection{A numerical instability of near-isothermal sonic flows}

\begin{figure*}
\centering
\begin{tabular}{cc}
\multicolumn{1}{l}{(a)} &
\multicolumn{1}{l}{(b)} \\
\epsfxsize=\textwidth\divide\epsfxsize by 2
\mbox{\epsfbox[40 400 515 715]{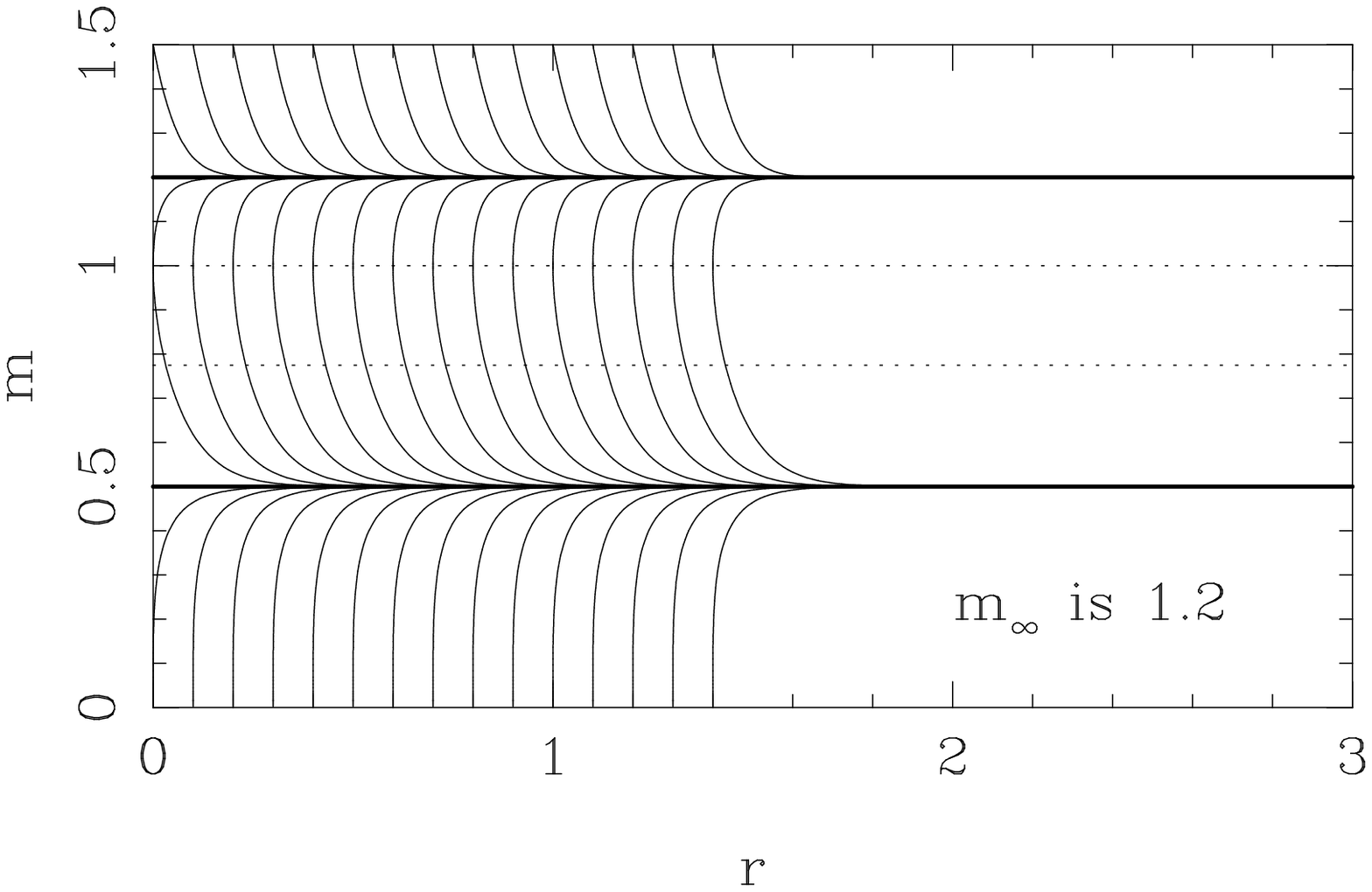}} &
\epsfxsize=\textwidth\divide\epsfxsize by 2
\mbox{\epsfbox[40 400 515 715]{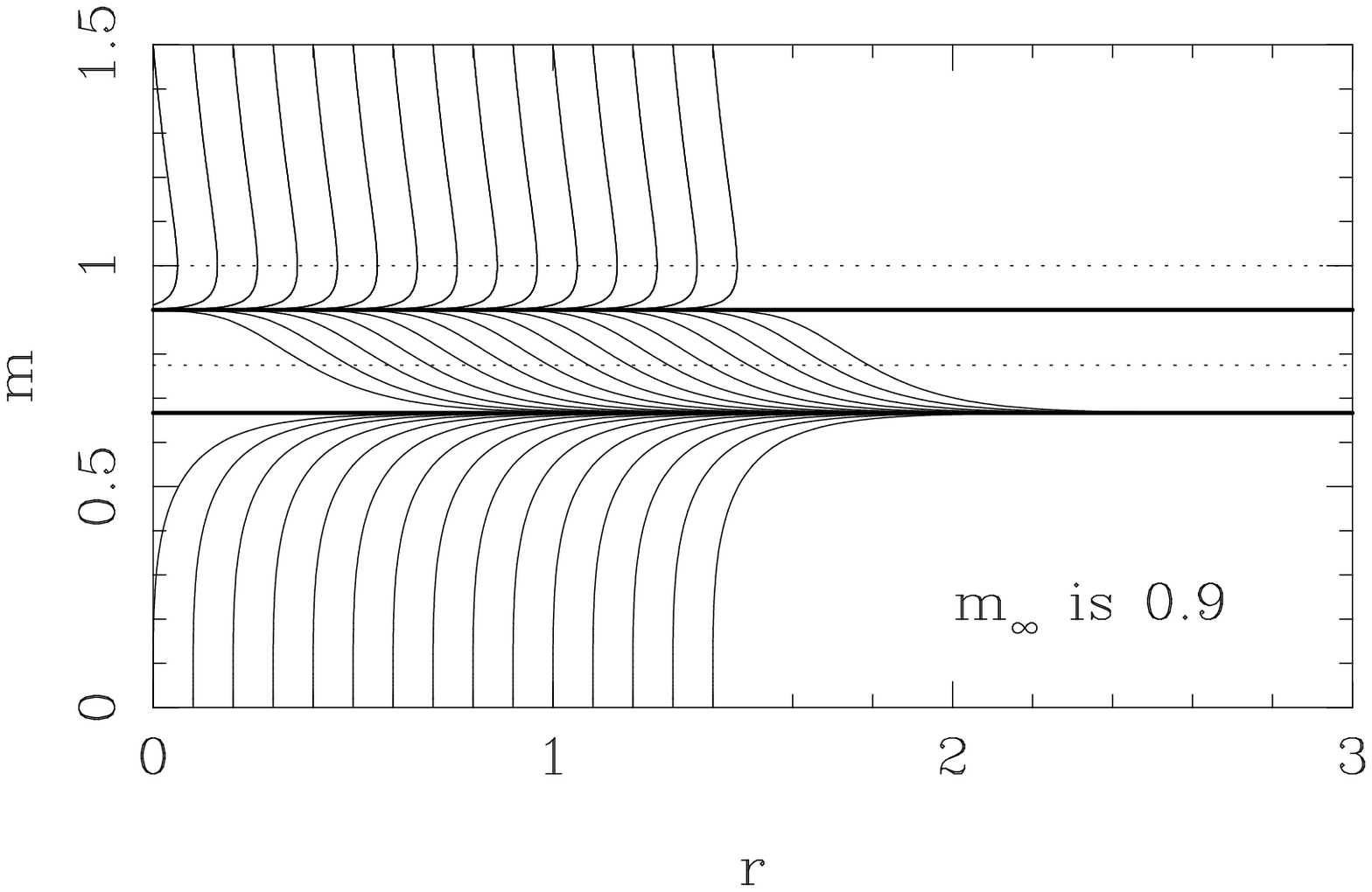}}
\end{tabular}
\caption{Steady solutions of flows with heating $Q\propto T|e-T$
for (a) $m_\infty=1.2$ and (b) $m_\infty = 0.9$.  The flow direction
is towards increasing $r$. The bold, horizontal lines are the
equilibrium solutions.  Dotted lines mark the isothermal and adiabatic
sound speeds.}
\label{f:trivial}
\end{figure*}
Using equations~\refeq{e:aux} to relate $m$ to $c$, we can rearrange
equation~\refeq{e:mfront} into the suggestive form:
\begin{equation}
v \rdif{\,\ln(c^2)}{r} =
	{1-\gamma m^2\over 1-m^2}(\gamma-1)Q/c^2,\label{e:eqm}
\end{equation}
(\cf\ also equation~(15) of Williams, Hartquist \& Dyson, 1995).  If
the level of ionization is weakly varying, then $c^2 \propto T$.
For substantially subsonic flows ($m\ll 1$) and for supersonic flows
($m > 1$) the solution will tend to relax to a stable equilibrium as
we integrate in the downstream direction.  For our system of
equations, the flow reaches equilibrium at a temperature $T\simeq
7800\Kelv$ (\cf\ NA).

However, when the flow velocity is in the range between the isothermal
sound speed (\ie\ $m = 1/\sqrt\gamma$) and the adiabatic sound speed
($m=1$), the first factor on the r.h.s. of equation~\refeq{e:eqm} is
negative: thermally stable equilibria appear to become unstable (when
the flow is integrated downstream) and vice versa (so long as the
dynamical time, $(\rdif{v}{r})^{-1}$, is much longer than the thermal
timescale, $\sim c^2/Q$).

This behaviour is readily understood in terms of the formation of weak
`isothermal shocks' in the flow.  If the flow were {\it strictly}\/
isothermal, discontinuous shocks could form with upstream Mach numbers
in the unstable range (\cf\ Figure~\ref{f:onetr}).  For finite
cooling, discontinuous solutions no longer exist.  The instability of
the equilibrium solution nevertheless allows the flow to decrease its
Mach number through a {\it resolved}\/ transition at any point, and
since the flow properties upstream and downstream of this transition
satisfy the isothermal Rankine-Hugoniot relations, we refer to these
transitions as resolved isothermal shocks.

We illustrate this in Figure~\ref{f:trivial} for the simpler case of a
steady plane flow with no ionization, using a simple heating function
$Q\propto (T|e-T)$ in equation~\refeq{e:mfront}, where $T|e$ is the
equilibrium temperature.  The plot shows rest frame solutions, with
the integration performed downstream, towards increasing $r$.  The
solutions are independent of $r$, but a range of solutions are shown
arbitrarily offset in this direction to indicate the morphology of
solutions expected in more complicated situations.

Note that in this model, $Q$ is independent of $\rho$.  This cooling
law is stable to equilibrium perturbations, and also marginally stable
to the instabilities of algebraic power within cooling zones
\cite{field65,fallr85}.  Since the peak temperatures found within the
isothermal shocks are less than 3 per cent above the equilibrium
temperature, we can safely ignore this marginal instability; in any
case, the results are similar for $Q \propto (T|e^n-T^n)$ with $n>1$.

We first show a case for which both the sub- and the supersonic
solutions to the flow equations integrate stably, with $m_\infty =
1.2$ (or, equivalently, $m_\infty = 0.5$).  In
Figure~\ref{f:trivial}a, the bold, horizontal lines are the
equilibrium solutions; thinner solid lines show solutions integrated
downstream (to the right) from an initial non-equilibrium state.  The
isothermal and adiabatic sound speeds are shown as dotted lines.  As
we integrate downstream, sub- (super-) sonic initial conditions relax
to the sub- (super-) sonic equilibrium solution.

Figure~\ref{f:trivial}b is for $m_\infty = 0.9$ (or, equivalently,
$m_\infty \simeq 0.67$).  Here, we see that the equilibrium solution
$m = 0.9$ is apparently unstable as we integrate downstream (left to
right), with solutions diverging to large $m$ and towards the $m
\simeq 0.67$ equilibrium.  These latter solutions are, in effect, weak
shocks.  If we consider a Galilean transformation of the system, it is
apparent that the instability found is entirely a result of the
steady-state assumption: $m = 0.9$ is a perfectly reasonable
equilibrium solution, as is confirmed by time-dependent simulations of
such flows.  For some downstream conditions, however, it will be
necessary for a weak shock to form in the flow at some (in the present
case, arbitrary) position.  The instability demonstrated here allows
this to occur in continuous solutions, just as discontinuous adiabatic
Rankine-Hugoniot shocks can form at arbitrary positions in flows
initially above the adiabatic sound speed.

Since downstream integration of these ODEs for initial conditions
perturbed from equilibrium confirms the existence of rapidly divergent
solutions for flows between the isothermal and adiabatic sound speeds,
we conclude that the solutions found by Newman \&
Axford~\shortcite{newa68} in this range of upstream Mach numbers
inadvertently included these weak isothermal shocks.  In the next
section, we present the additional transonic solutions this led them
to disregard.

\subsection{Structure of recombination fronts}

\begin{figure}
\centering
\epsfxsize=\textwidth\divide\epsfxsize by 3
\mbox{\epsfbox[96 37 480 675]{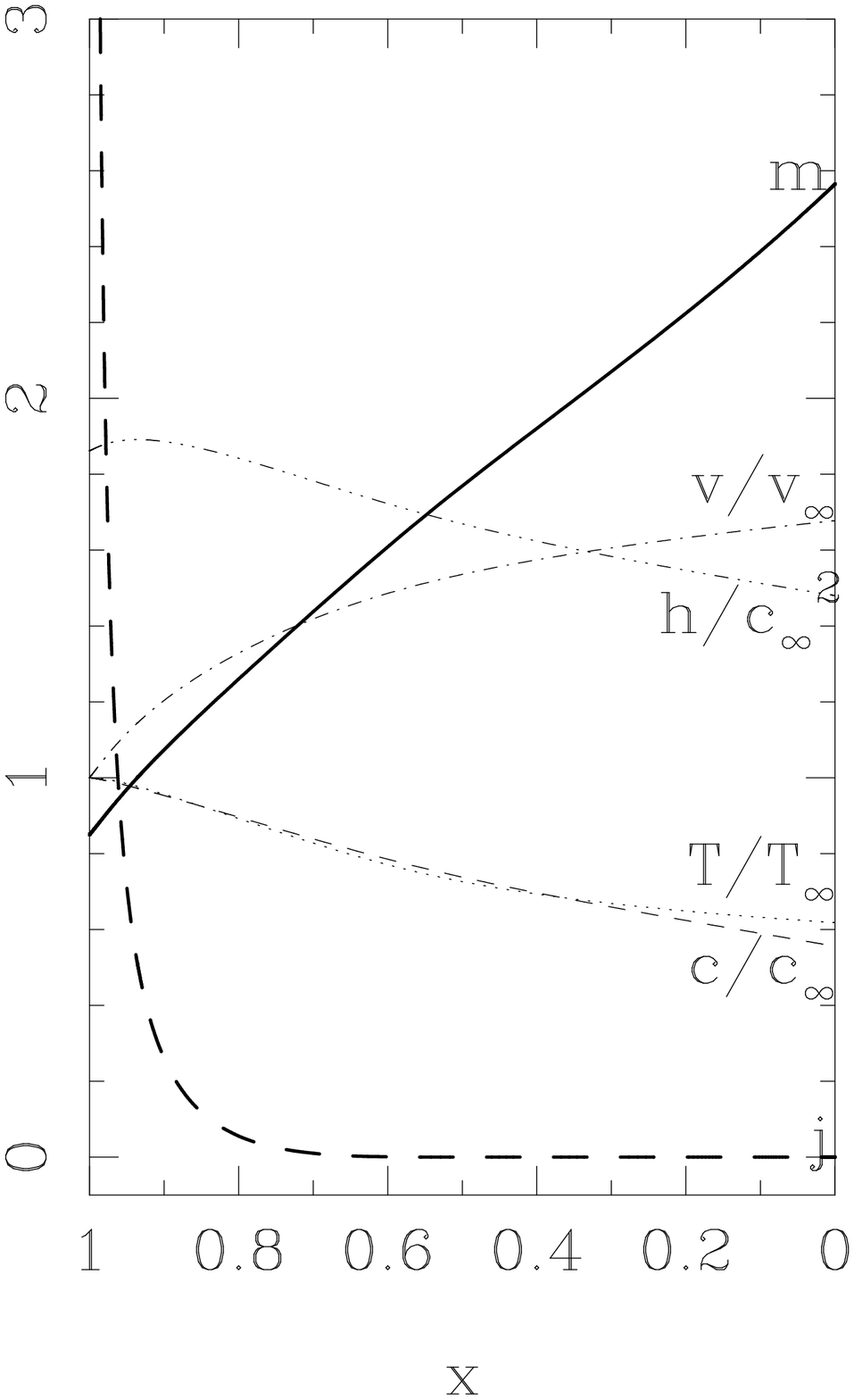}}
\caption{Transonic solution for an RF with
Mach number $m_\infty = 0.85$ far upstream in the ionized gas.  The
Mach number $m$, dimensionless flux $j$, sound speed $c$, temperature
$T$, flow velocity $v$ and specific enthalpy $h$ are plotted against
ionization fraction $x$.  This solution here has $j|{son} = 0.51$ and
$x|{son} = 0.936$, and the Mach number of the \HI\ flow far downstream
is $m_1 = 2.56$, corresponding to a temperature there of $4800\Kelv$.}
\label{f:transp}
\end{figure}
\begin{figure*}
\centering
\epsfclipon
\begin{tabular}{cc}
\multicolumn{1}{l}{(a)} &
\multicolumn{1}{l}{(b)} \\
\epsfxsize=\textwidth\divide\epsfxsize by 2
\mbox{\epsfbox[194 25 395 337]{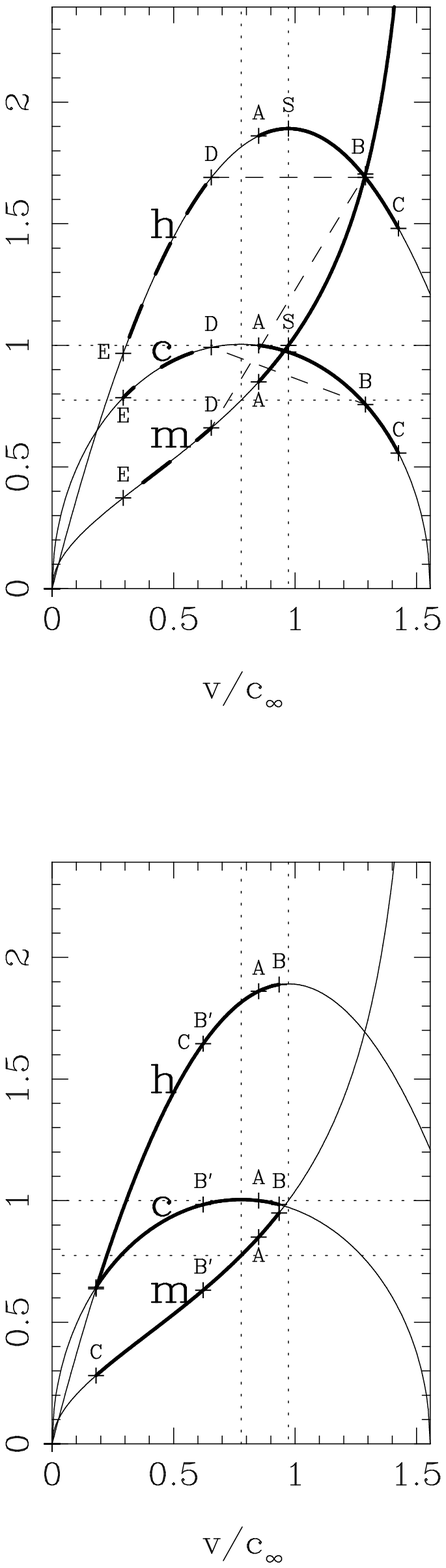}} &
\epsfclipon
\epsfxsize=\textwidth\divide\epsfxsize by 2
\mbox{\epsfbox[194 402 395 714]{paths.ps}}
\end{tabular}
\caption{Example solutions for $m_\infty = 0.85$.  The enthalpy ($h$, in
units of $c_\infty^2$), sound speed ($c$ in units of $c_\infty$) and
adiabatic Mach number ($m$) are plotted against the flow velocity ($u$
in units of $c_\infty$).  Vertical dotted lines mark the maximum of
$c$, where the flow is at the isothermal sound speed, and the maximum
of $h$, where the flow is at the adiabatic sound speed; horizontal
dotted lines mark the values of $m$ at these speeds.  In case (a) an
`isothermal shock' is formed ahead of the sonic point (the flow
travels $\rm A\to B$ then passes through the isothermal shock $\rm
B\to A \to B'$ and then recombines and cools to reach $\rm C$ in the
end).  Case (b) includes a sonic point (S) -- the solution with no
shock is $\rm A\to S\to B\to C$, but there are also solutions such as
$\rm A\to S\to B\to D\to E$ which include an adiabatic shock ($\rm
B\to D$, shown dashed).}
\label{f:paths}
\end{figure*}
\begin{figure*}
\centering
\epsfxsize=\textwidth
\mbox{\epsfbox[36 400 560 750]{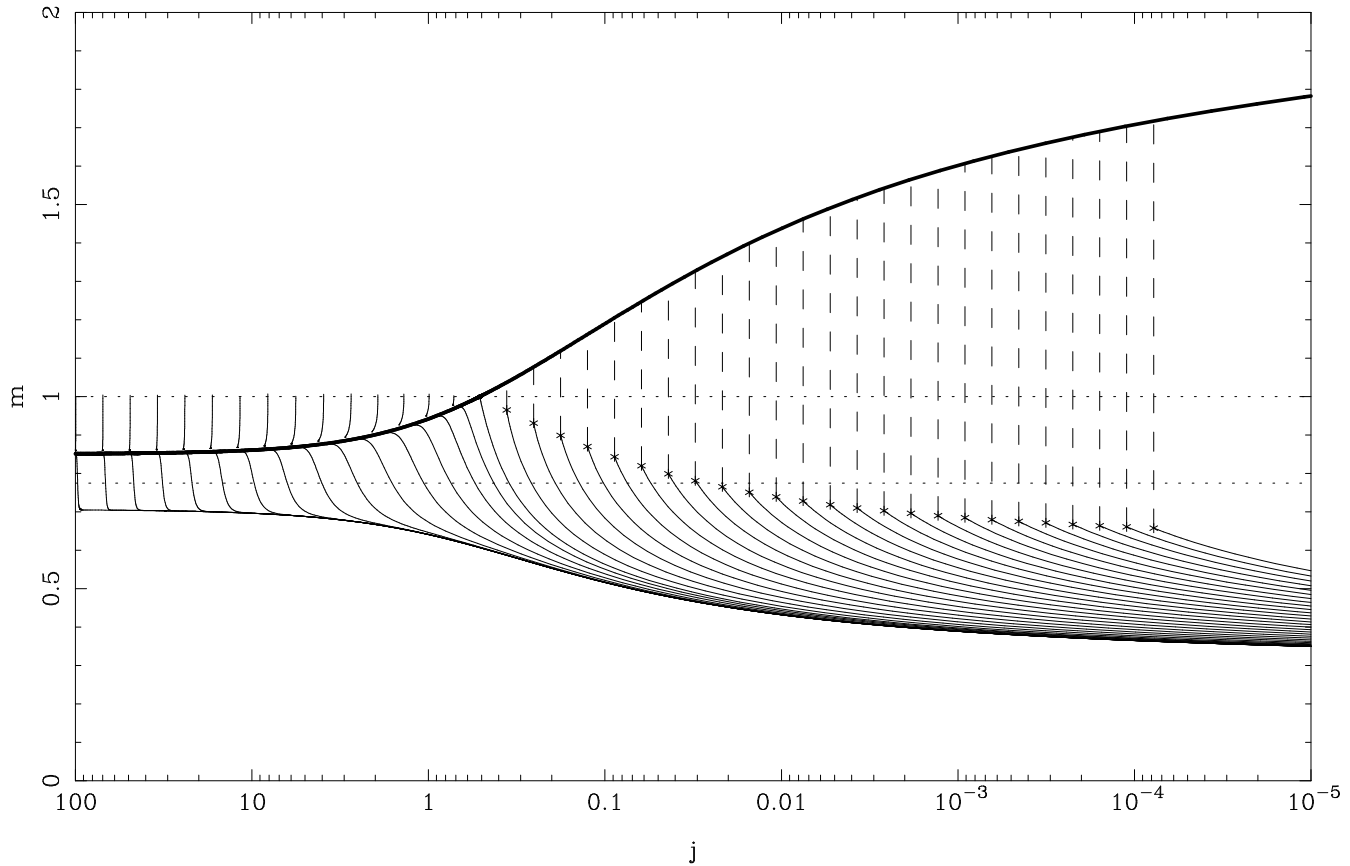}}
\caption{Steady solutions for an RF with
Mach number $m_\infty = 0.85$ far upstream (at high ionizing flux, to
the left of this plot) in the ionized gas. The bold curve is the
transonic solution.  A sample of the solutions divergent from the
transonic case are shown for initial points upstream of the sonic
point.  Downstream of the sonic point, solutions are shown, perturbed
by an initial steady adiabatic shock (shown dashed).  Horizontal
dotted lines mark the isothermal and adiabatic sound speeds.  For the
subsonic downstream solutions, final temperatures vary between
$6500\Kelv$ (for flows which shock far upstream in the \HII\ region)
and $14000\Kelv$ (where the gas shocks far downstream in the \HI\
region).}
\label{f:onetr}
\end{figure*}

\begin{figure}
\centering
\epsfxsize=\textwidth\divide\epsfxsize by 2
\mbox{\epsfbox[43 138 549 615]{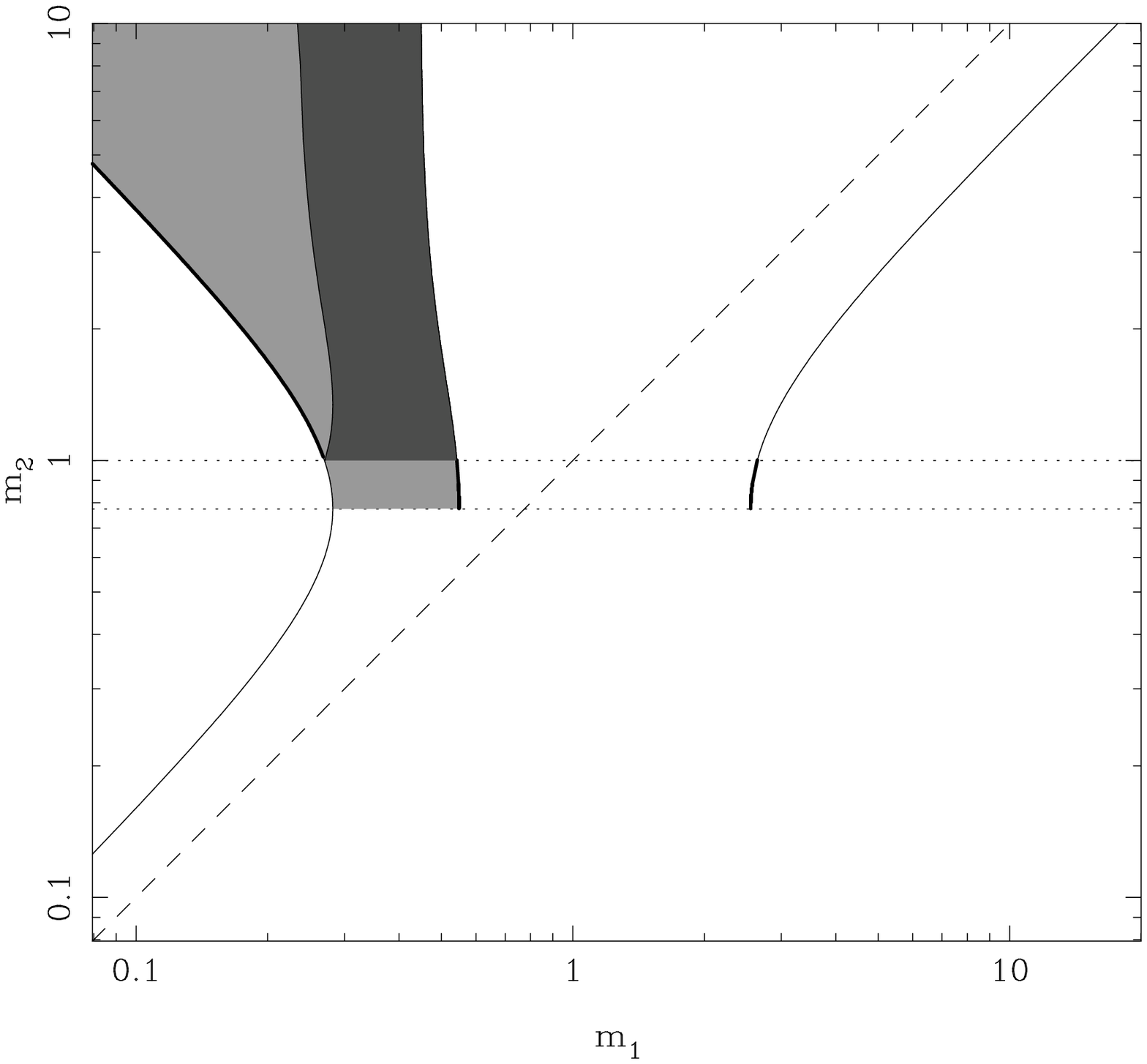}}
\caption[.]{Transition diagram for RF (\cf\ Figure~5 of
NA).  Here $m_2$ is the adiabatic Mach number far upstream in the
\protect\HII\ flow, $m_1$ is the adiabatic Mach number far downstream
in the \protect\HI\ flow.  The dashed line is $m_1=m_2$; dotted lines
mark where the upstream flow is at the isothermal sound speed, $m_2 =
1/\sqrt{\gamma}$, and at the adiabatic sound speed $m_2 = 1$.  The
narrow lines and dark gray area are the same as those in Newman \&
Axford (although note that they use isothermal, rather than adiabatic,
Mach numbers).  The bold lines and lighter shaded area are also
allowed. In particular, note the short section of the $m_1>1$ curve
with between the horizontal dotted lines: this section is the locus of
transonic RF\@.}
\label{f:transitions}
\end{figure}

The presence of the instability discussed in the previous section in
the downstream integration of near isothermal flows implies that, when
the upstream flow is between the isothermal and adiabatic sound
speeds, the initial conditions $m \to m_\infty$, $x \to 1$ at $j \to
\infty$ used by Newman \& Axford {\it do not fully specify}\/ the flow
solutions.  In this section, we use this freedom to find an important
class of RF overlooked by NA -- transonic fronts.  In what follows, we
return to the full model equations~(\ref{es:thinfront}a-c) for the
structure of RF.

In looking for transonic solutions, where $Q=0$ and $v(\rdif{Q}{r}) < 0$
at $m=1$ (so the flow can pass smoothly through the adiabatic sound
speed), we have formulated a two-point boundary problem (with
conditions applied both far upstream in the \HII\ flow and at the
sonic point).  We have indicated that for initial conditions
$1/\sqrt{\gamma} < m < 1$, the Mach equation is unstable when
integrated downstream.  However, the ionization equation is highly
unstable when integrated in the opposite, upstream direction.

The vigorousness of these instabilities precludes the use of shooting
methods.  We therefore employed an iterative procedure to find the
flow solution between $j\to\infty$ and the adiabatic sonic point
(outside this region, downstream integration proceeds smoothly).  An
approximation to the ionization structure was determined by downstream
integration.  Using this approximation to the ionization structure,
the equation for the Mach number was integrated upstream.  This gave
an updated model for the velocity and sound speed distributions, which
were then used to derive an updated ionization structure: this
procedure was iterated to convergence.

The integrations were performed with a stiff ODE integration
package \cite{cohh94}, using cubic spline interpolation of the
results.  The singular equations for $\rdif{m}{j}$ and $\rdif{x}{j}$
were split into non-singular equations in a dummy independent variable
$s$ (\ie\ for $\rdif{m}{s}$, $\rdif{x}{s}$ and $\rdif{j}{s}$).

In Figure~\ref{f:transp} we show one such transonic solution, for
upstream Mach number $m_\infty=0.85$.  The solid bold curve is the
Mach number of the transonic solution, plotted as a function of the
ionization fraction, $x$.  Also shown in this figure are the
distributions of dimensionless ionizing flux $j$, sound speed $c$,
temperature $T$, velocity $v$ and stagnation enthalpy, $h =
c^2/(\gamma-1)+v^2/2$, through the front.  Note that all these curves
are smooth, and that $h$ has a maximum value when $m=1$, as expect
(\cf\ Figure~\ref{f:paths}).

In Figure~\ref{f:paths}, we show the various classes of solution
perturbed from the transonic case.  The three curves on the plots show
the adiabatic Mach number, $m$, sound speed, $c$, and enthalpy $h$ as
a function of flow velocity, $v$.  In Figure~\ref{f:paths}a, a
solution with a weak isothermal shock is shown.  From the initial
point A, the flow cools and accelerates.  However, an isothermal shock
is triggered at point B, and the flow then heats again and slows.  As
the shock is resolved, the solution remains on the integral curves at
all times -- the part of the curve between points B and B$'$ might,
however, be considered to represent the isothermal shock.  Beyond the
shock, in the region $\rm B'\to C$, the flow decelerates and cools
further as the gas recombines.

Figure~\ref{f:paths}b illustrates a transonic solution, $\rm A\to S\to
B\to C$.  Here, the flow has been chosen to accelerate smoothly
through sonic point S, at the maximum of the enthalpy $h$.  Beyond the
sonic point, the flow accelerates, cools and recombines to reach point
C when it has become fully neutral.  Also illustrated is the solution
which results when an adiabatic shock occurs at point B, taking the
flow discontinuously to D.  Beyond D, the flow behaves in a similar
fashion to that illustrated for the isothermal shock case in
Figure~\ref{f:paths}a, except that it does not reach quite such a low
temperature as it approaches infinity.

In Figure~\ref{f:onetr} we plot the flow Mach number, $m$, for a wide
range of these solutions against the dimensionless ionizing flux, $j$.
Upstream of the sonic point, both the physical solution (which
continues subsonically to $j\to 0$) and the unphysical solution (which
tends to $m=1$ with infinite gradient) are shown.  At large $j$, these
solutions diverge very sharply, underlining the difficulty in finding
transonic solutions by shooting methods.  Beyond the sonic point,
solutions are shown for initial conditions found downstream of steady
adiabatic shocks (shown dashed) which are now allowed.  These shocked
solutions form a one-parameter family in combination with the physical
solutions which diverge from the transonic solution upstream of the
sonic point: as the stepping-off point of these solutions moves
progressively downstream, the maximum in Mach number of the divergent
solutions becomes a cusp at the sonic point, and then a discontinuity
(\ie\ shock).  A range of final Mach numbers results from this range
of solutions.

In Figure~\ref{f:transitions} we show the transition diagram from Mach
number $m_2$ in the \HII\ flow (at the equilibrium temperature) far
upstream to Mach number $m_1$ in the downstream \HI\ flow.  In this
diagram, we plot both the results we obtained following Newman \&
Axford, and the extra solutions found in this work.  The short segment
of the upper right hand branch of solutions shown bold, between dotted
horizontal lines at the isothermal and adiabatic sound speeds, give
the transonic solutions.  Additional regions of subsonic solutions to
the top left result both from shocks downstream of the sonic point in
transonic solutions, and from the use of isothermal, rather than
adiabatic, shock relations for shocks forming in the \HII\ region.

To interpret this diagram, it is useful to remember that because the
upstream flow is assumed to be at the fixed equilibrium temperature,
the curves plotted are for a {\it family}\/ of ODEs, parameterised by
$\Pi/\Phi$.  Isothermal shock conditions $m_+ m_- = 1/\gamma$ relate
the values of $m_2$ where the ODEs are identical.  The sudden
appearance of extra solutions with $m_1 > 1$ at $m_2 =
1/\sqrt{\gamma}$ is allowed because this is the singular value where
$\Pi/\Phi$ is a minimum.

\section{Conclusions}

In this paper, we have completed the solution space found by Newman \&
Axford~\shortcite{newa68} for recombination front flows.  Careful
treatment of the numerical integration for flow speeds between the
isothermal and adiabatic sound speeds reveals an extra class of
transonic solutions overlooked by these authors.

Close to the sonic point, the form of these solutions is similar to
those found for any steady transonic flow, for instance a
Parker~\shortcite{park58} wind.  The steady, transonic solution can in
general be found by integrating away from the sonic point.  However,
inclusion of ionization equilibrium means that we must treat a
two-point boundary value problem.  The strength of the instability
discussed above adds significantly to the usual difficulties
integrating the Mach number equation towards a nodal solution,
rendering shooting methods impractical and allowing transonic
solutions to be easily overlooked.  We overcome this difficulty with
an iterative scheme, integrating separate equations in the upstream
and downstream directions.

The instability in downstream integration adds a free parameter to the
solution, resolving the conceptual difficulty of how, as the initial
Mach number of our solution increases through isothermal Mach 1, the
flow solution changes from an initial-value to a two-point boundary
problem.  Moreover, these solutions give some insight into the
apparent paradox that, for finite cooling, the singularity in steady
flow equations is at the adiabatic sound speed, whereas in a strictly
isothermal flow it is at the isothermal sound speed.  We find that,
where thermal timescales are short in an accelerating flow, a `wide'
range of solutions near the adiabatic sound speed derive from a `few'
at the isothermal sound speed: it is apparent that in the limit, these
`few' solutions can become the singular isothermal transonic solution.

In the theory of multifluid MHD shocks, smooth transonic solutions are
also found for strong cooling \cite{cher87,flowp87,robed90}.  In these
shocks, the solution adapts to pass through a downstream (saddle type)
sonic point using the freedom of manoeuvre allowed by an upstream
viscous subshock (J-type) or nodal sonic point (C$^\star$-type),
rather than as discussed here.

There has been little discussion of RF since the work of NA, since in
the winds of main sequence stars they will occur, if at all, at large
radii, and thus at densities and temperatures so low that they will be
broadened beyond recognition by advection.  The interstellar medium in
the vicinity of strongly ionizing stars does not generally reach an
equilibrium state in the stellar lifetime -- an IF is still being
driven out into the surrounding medium when the star dies.  However,
in recent work we have discussed how a near-equilibrium state may
found in the youngest of these stars, observed as ultra compact \HII\
regions \cite{dyso94,wild94,dywr95}.  Mass loading in the near
vicinity of the young stars traps the IF at a scale of $0.1\parsec$,
turning it into a RF through which gas passes at high density.  In
some objects, the mass-loaded stellar wind may remain supersonic at
all radii -- however, if the mass loading is strong enough the flow
will shock.  Detailed hydrodynamical simulations (Williams, Redman \&
Dyson, in preparation) confirm that transonic RF are frequently a
component of the equilibrium flow.  Such regions may be a significant
source for partially ionized gas in the interstellar medium.

We note briefly that the discussion here also applies to the structure
of \HII\ region IF as they undergo the transition from R-type to
D-type \cite{maso80}.  A shock first forms, as a (quasi-) steady part
of the IF structure, at the minimum of the adiabatic Mach number in
the R-type front when this reaches unity; initially, the post-shock
flow is in the unstable range.  In this region, however, the heating
and cooling in the transonic flow are rather weak.  More
significantly, as the resultant strong D-type front weakens, the
velocity relative to the front downstream in the \HII\ region will
eventually reach the unstable range.  In this case, there is a
significant chance that solutions will be missed in steady-state
integrations.

\section*{Acknowledgements}

We thank Robert Cannon, Sam Falle, Franz Kahn and Matt Redman for
helpful discussions.  The referee's thoughtful report improved the
exposition and made us aware of analogous results in the literature on
MHD shocks.  This work was supported by PPARC through the Rolling
Grant to the Astronomy Group at Manchester (RJRW).

\bsp
\label{lastpage}

\begin{thebibliography}{}

\bibitem[\protect\citename{Axford }1961]{axfo61}
Axford W.I., 1961, Phil.\ Trans.\ R. Soc.\ London, A, 253, 301

\bibitem[\protect\citename{Beltrametti, Tenorio-Tagle \& Yorke }1982]{bety82}
Beltrametti M., Tenorio-Tagle G., Yorke H.W., 1982, A\&A, 112, 1

\bibitem[\protect\citename{Chernoff }1987]{cher87}
Chernoff D.F., 1987, ApJ, 312, 143

\bibitem[\protect\citename{Churchwell }1990]{chur90}
Churchwell E., 1990, A\&AR, 2, 79

\bibitem[\protect\citename{Cohen \& Hindmarsh }1994]{cohh94}
Cohen S.D., Hindmarsh A.C., 1994, CVODE User Guide, netlib archive

\bibitem[\protect\citename{Dyson \& Williams }1980]{dysw80}
Dyson J.E., Williams D.A., 1980, The Physics of the Interstellar Medium,
Manchester University Press, Manchester

\bibitem[\protect\citename{Dyson }1994]{dyso94}
Dyson J.E., 1994, in Ray T.P., Beckwith S.V., ed, Lecture Notes in
Physics 431: Star Formation Techniques in Infrared and mm-wave
Astronomy, Springer-Verlag, Berlin, p.\ 93

\bibitem[\protect\citename{Dyson, Williams \& Redman }1995]{dywr95}
Dyson J.E., Williams R.J.R., Redman M.P., 1995, MNRAS in press

\bibitem[\protect\citename{Fall \& Rees }1985]{fallr85}
Fall S.M., Rees M.J., 1985, ApJ 298, 18

\bibitem[\protect\citename{Field }1965]{field65}
Field G.B., 1965, ApJ, 142, 531

\bibitem[\protect\citename{Flower \& Pineau des For\^{e}ts }1987]{flowp87}
Flower D.R., Pineau des For\^{e}ts G., 1987, MNRAS, 224, 403

\bibitem[\protect\citename{Frank, Noriega-Crespo \& Balick }1992]{frnb92}
Frank A., Noriega-Crespo A., Balick B., 1992, AJ 104, 841

\bibitem[\protect\citename{Goldsworthy }1961]{gold61}
Goldsworthy F.A., 1961, Phil.\ Trans.\ R. Soc.\ London, A, 253, 277

\bibitem[\protect\citename{Kahn }1954]{kahn54}
Kahn F.D., 1954, Bull.\ Astr.\ Inst.\ Netherlands, 12, 187

\bibitem[\protect\citename{Mason }1977]{maso77}
Mason D.J., 1977, Ph.D. Thesis, University of Manchester

\bibitem[\protect\citename{Mason }1980]{maso80}
Mason D.J., 1980, A\&A, 92, 117

\bibitem[\protect\citename{Mallik }1975]{mall75}
Mallik D.C.V., 1975, ApJ, 197, 355

\bibitem[NA]{newa68}
Newman R.C., Axford W.I., 1968, ApJ, 151, 114

\bibitem[\protect\citename{Osterbrock }1989]{oste89}
Osterbrock D.E., 1989, Astrophysics of Gaseous Nebulae and Active
Galactic Nuclei, University Science Books, Mill Valley, CA

\bibitem[\protect\citename{Parker }1958]{park58}
Parker E.N., 1958, ApJ, 128, 664

\bibitem[\protect\citename{Roberge \& Draine }1990]{robed90}
Roberge W.G., Draine B.T., 1990, ApJ, 350, 700

\bibitem[\protect\citename{Williams \& Dyson }1994]{wild94}
Williams R.J.R., Dyson J.E., 1994, MNRAS, 270, L52

\end{thebibliography}
\end{document}